\documentclass[twocolumn,prl,aps,showpacs,superscriptaddress]{revtex4}
\usepackage{epsfig}
\usepackage{bm,hyperref}
\usepackage{amssymb}
\def\r{\bm{R}}

\def\h{\mathcal{H}}

\def\be{\begin{equation}}
\def\ee{\end{equation}}
\def\ba{\begin{eqnarray}}
\def\ea{\end{eqnarray}}

\begin{document}
\title{Adiabatic Geometric Phase for a General Quantum State}
\author{Biao Wu}
\affiliation{Condensed Matter Sciences Division, Oak Ridge National
Laboratory, Oak Ridge, Tennessee 37831, USA}
\affiliation{Institute of Physics, Chinese Academy of Sciences, 
Beijing 100080, China}
\author{Jie Liu}
\affiliation{Institute of Applied Physics and Computational Mathematics, 
P.O. Box 8009, Beijing 100088, China}
\author{Qian Niu}
\affiliation{Department of Physics, The University of Texas, Austin, 
Texas 78712, USA}
\date{March 29th, 2004}

\begin{abstract}
A geometric phase is found for a general quantum state 
that undergoes adiabatic evolution. For the case of eigenstates,
it reduces to the original Berry's phase. Such a phase is 
applicable in both linear and nonlinear quantum systems. Furthermore, 
this new phase is related to Hannay's angles as we find that these angles, 
a classical concept, can arise naturally in quantum systems. 
The results are demonstrated  with a two-level model.
\end{abstract}
\maketitle
\begin{figure}[!ht]
\vskip10pt
\center{\includegraphics[width=6.0cm]{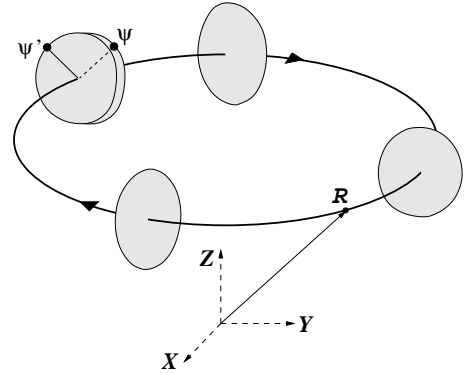}}
\caption{Adiabatic evolution of a quantum state along a closed path in
the parameter space of $\r$. In general, the quantum state does not come
back at the end of the evolution, that is, the difference between 
the start state $\psi$ and the end state $\psi^\prime$ is more than a phase.
The disks represent the dynamical evolution of a quantum state at a given
$\r$; their different shapes symbolically illustrate the change of the system 
Hamiltonian with $\r$. The areas of these disks remain the same,
reflecting that the actions $\bm{I}$ are conserved.}
\label{fig:torus}
\end{figure}

Consider a quantum system that depends on some external parameters $\r$.
We are interested in the evolution of its quantum state when the parameters
$\r$ change slowly along a closed path (see Fig.\ref{fig:torus}).
For an eigenstate, such an adiabatic evolution accumulates a 
geometric phase, which reflects the system geometry with the parameter 
space $\r$. It is now known as Berry's phase\cite{berry}. Although the 
same geometry is also embedded in the adiabatic evolution of a general 
quantum state, including non-eigenstate, it is not clear how a geometric 
phase can be defined for a general quantum state to extract the system 
geometry.

In this Letter we introduce a geometric phase for a 
general quantum state. In the case of eigenstates, this phase can be 
reduced naturally to the original Berry's phase. From this reduction, 
one gains some fresh perspective on the familiar concept of Berry's phase.

The geometric nature of this new phase is fully explored. Similar to 
Berry's phase, it can be regarded as the geometric part of
a total phase associated with the adiabatic evolution. In another
perspective, it can also be regarded as a part of an
Aharonov-Anandan (AA) phase\cite{aa,chiao}. Moreover, the new phase is 
found to be related to Hannay's angles\cite{hannay} in a 
derivative form. Although Hannay's angles have long been considered 
as the counterpart of Berry's phases in classical mechanics, we find 
that they can be defined naturally in quantum systems, 
and be calculated from the new phase. In this sense, this relation unifies 
two very different concepts, Berry's phase and Hannay's angles.

Interestingly, this new geometric phase is also applicable in
nonlinear quantum systems. Unlike in linear quantum systems where one may
understand the adiabatic evolution of a non-eigenstates in terms of 
eigenstates\cite{anandan}, it is impossible to do the same in nonlinear 
quantum systems due to the lack of the superposition principle. Therefore,
the introduction of this new phase provides a unique and powerful tool to 
study the adiabatic evolution of a general state in nonlinear 
quantum systems. Bose-Einstein condensates of dilute atomic gases are an 
excellent example of nonlinear quantum systems\cite{bec_rmp,leggett}.

In the following discussions, we will proceed with nonlinear quantum systems 
since linear quantum systems can be regarded as their special cases.

We consider an $N$-level quantum system governed
by a general nonlinear Schr\"odinger equation ($\hbar=1$),
\be
\label{eq:nls}
i\frac{d}{dt}|\psi\rangle=H(\psi_j^*,\psi_j;\r)|\psi\rangle\,.
\ee  
Here $|\psi\rangle=(\psi_1,\psi_2,\cdots,\psi_N)$ is the wave function with
$\psi_j$ being its $j$th component over an orthonormal basis;
the vector $\r$ represents all the system parameters 
subject to adiabatic change. We assume that the system  is gauge invariant 
since it is the case of most physical interest. 
When $H$ is independent of $\psi^*_j$ and $\psi_j$, Eq. (\ref{eq:nls}) is
the usual linear Schr\"odinger equation.

It is well known that the quantum system governed by Eq.(\ref{eq:nls}) 
mathematically has a canonical classical Hamiltonian structure (e.g.,
see Refs.\cite{weinberg,heslot}). That is, one can find a Hamiltonian
 $\mathcal{H}(\psi_j^*,\psi_j;\r)$
such that Eq. (\ref{eq:nls}) is a set of equations of motion 
when $\mathcal{H}$ serves the classical Hamiltonian with Poisson brackets 
$\{\psi_j^*,\psi_k\}=i\delta_{jk}$. When the Hamiltonian is bilinear, 
$\h=\langle\psi|H|\psi\rangle$, the system is linear.
With this Hamiltonian structure, the system of Eq.(\ref{eq:nls}) 
can be classified into integrable or non-integrable in the classical sense. 
All linear quantum systems  are integrable classically.

We focus on the case that Eq.(\ref{eq:nls}) 
is integrable. In this case, the system at a given $\r$ has $N$ constants 
of motion, $\bm{I}=\{I_1,I_2,\cdots,I_N\}$. They are called actions, whose
conjugate variables are angles $\bm{\theta}=\{\theta_1,
\theta_2,\cdots,\theta_N\}$, which change with 
time linearly as $\bm{\theta}=\bm{\theta}_0+\bm{\omega}t$, where 
$\omega_j=\partial\mathcal{H}/\partial I_j$.
Therefore, at a given $\r$, the wave function can be expressed as a 
function of $\{\bm{I},\bm{\theta}\}$,  
$|\psi\rangle=|\psi(\bm{I},\bm{\theta};\r)\rangle$.  Such a parameterization
of the wave function in terms of $\{\bm{I},\bm{\theta}\}$ is also very 
convenient for slowly changing $\r$. In such an adiabatic evolution, 
the actions $\bm{I}$ are still conserved\cite{arnold} while the angles change as 
$\bm{\theta}(t)=\bm{\theta}_0+\int\bm{\omega}(\r)dt+\bm{\alpha}(t)$,
where $\bm{\alpha}$ are Hannay's angles\cite{hannay}.

With the above observations, we are now mathematically ready to
introduce a geometric phase for the adiabatic evolution of a general quantum 
state $|\psi\rangle$, where the parameters $\r$ change slowly along a closed 
path $\mathcal O$, as shown in Fig.\ref{fig:torus}. One feature 
immediately stands out: Except eigenstates, the quantum state generally 
does not come back to the original state at the end of evolution even 
though the system Hamiltonian recovers its original form. Moreover, 
the difference between the initial state and the ending state depends 
on the choice of the initial state. These features make it difficult 
to define Berry's phase and its various generalizations for this 
evolution\cite{aa,chiao,samuel}.

To circumvent the obstacle, we use an averaging technique as in the 
definition of Hannay's angles\cite{hannay}. For a quantum states 
$|\psi\rangle=|\psi(\bm{I},\bm{\theta};\r)\rangle$ adiabatically evolving along 
the path $\mathcal O$, we introduce a geometric  phase $\gamma$ as
\be
\label{eq:gb}
\gamma({\mathcal O})=\frac{\oint d\bm{\theta}}{(2\pi)^N}\oint_{\mathcal O} d\bm{R}
\langle\psi(\bm{I},\bm{\theta};\r)|i\frac{\partial}{\partial \bm{R}}
|\psi(\bm{I},\bm{\theta};\r)\rangle\,,
\ee
where the integration over $\bm{\theta}$ is to average over all possible 
quantum states with the same actions $\bm{I}$ at a given $\r$. 
We emphasize that the integration over $\r$ in Eq.(\ref{eq:gb}) is done 
for a fixed $\bm{\theta}$. This new geometric phase $\gamma$ is
the same for all the quantum states that has the same actions $\bm{I}$;
therefore, it may also be regarded as a geometric phase for an 
adiabatically evolving manifold. To illustrate and further 
explore this new phase $\gamma$, we start with some special cases 
and end with a demonstration with a nonlinear two-level model.

For a quantum system starting at an eigenstate $|E_n(\bm{R})\rangle$ defined by 
\be
H(\r)|E_n(\r)\rangle=E_n(\r)|E_n(\r)\rangle, 
\ee
it evolves dynamically as $|\psi(t)\rangle=e^{-i\beta(t)}|E_n(\r)\rangle$ when $\r$
changes adiabatically. In this special case, there is only one action,
the norm $I=\langle E_n(\r)|E_n(\r)\rangle$, whose corresponding angle 
is the phase $\theta=\beta$. Plugging this $|\psi\rangle$ into 
Eq.(\ref{eq:gb}) and noticing that the partial derivative over $\r$ does 
not act on $\theta$ and $I$, we obtain the phase $\gamma$ for 
eigenstate $|E_n\rangle$,
\be
\label{eq:be}
\gamma_n({\mathcal O})
=\oint_{\mathcal O} d\bm{R}\langle E_n(\r)|
i\frac{\partial}{\partial \bm{R}}|E_n(\r)\rangle\,,
\ee
which recovers the original Berry's phase\cite{berry}. Note that 
Eq.(\ref{eq:be}) is valid in both linear and nonlinear quantum 
systems, indicating that the original definition of Berry's phase
can be directly borrowed for nonlinear quantum systems if
only eigenstates are considered.

We turn to a general quantum state in linear quantum 
systems. In this linear case, we can expand the evolving quantum state in 
terms of the eigenstates,
\be
\label{eq:linq}
|\psi(t)\rangle=\sum_{n=1}^N a_n(t) |E_n(\r)\rangle\,.
\ee 
According to the quantum adiabatic theorem\cite{messiah},
the occupation probabilities of different eigenstates  $|a_n|^2$ are adiabatic 
constants. In fact, they are actions
$I_n=|a_n|^2$ when the system is regarded mathematically as a 
classical Hamiltonian system; their corresponding angle variables
$\theta_n$ are the phases of $a_n$'s. With these in mind,
computation of Eq.(\ref{eq:gb}) with the state (\ref{eq:linq}) is 
straightforward. We find that the off-diagonal terms are zeros after 
the averaging, and the geometric phase $\gamma$ is
\be
\label{eq:lingb}
\gamma({\mathcal O})=\sum_{n=1}^N |a_n|^2\gamma_n({\mathcal O})\,,
\ee
where $\gamma_n$ is the Berry's phase of eigenstate $|E_n\rangle$. Therefore,
in linear quantum systems, the phase $\gamma$ is just a weighted 
summation of the Berry's phases of all the eigenstates involved.
Interestingly, this kind of weighted summation of Berry's phases has already
been applied  in calculating transverse fore on a quantized vortex\cite{ao}
and the anomalous Hall conductivity of ferromagnets\cite{yao}.

We have so far illustrated the geometric phase $\gamma$ for some 
simple examples and shown clearly how the phase $\gamma$ as defined 
in Eq.(\ref{eq:gb}) are related to the well-known Berry's phase. 
In the following, we are going to examine the new phase $\gamma$ in 
a general setting and derive for it a different expression.
These efforts reveal that the phase $\gamma$ can be regarded as a 
geometric part of the total phase of an adiabatic evolution, 
similar to  Berry's phase.

Imagine an adiabatic evolution of a general 
quantum state $|\psi\rangle$ along a close path $\mathcal O$.
Without worrying about whether it is geometric or not,
we can always introduce a phase for such an evolution,
\be
\label{eq:taa}
\beta=i\int_0^T dt \langle\psi|\frac{d}{dt}|\psi\rangle
=i\oint_{\mathcal O} \langle\psi|d|\psi\rangle \,,
\ee
where $T$ is the total evolution time. We call this phase the total phase
of the adiabatic evolution of $|\psi\rangle$. We expand the integrand as
\ba
\langle\psi(\bm{I},\bm{\theta};\r)|d|\psi(\bm{I},\bm{\theta};\r)\rangle
=\langle\psi|\frac{\partial}{\partial\bm{I}}|\psi\rangle\cdot d\bm{I}+\nonumber\\
+\langle\psi|\frac{\partial}{\partial\bm{\theta}}|\psi\rangle\cdot d\bm{\theta}
+\langle\psi|\frac{\partial}{\partial\r}|\psi\rangle\cdot d\r\,.
\ea
For an adiabatic evolution, the actions are conserved so
the term involving $\bm{I}$ is zero.
Plugging it back into Eq.(\ref{eq:taa}) and averaging it over the 
initial angles $\bm{\theta}_0$, we obtain 
\ba
\bar{\beta}&=&\oint\frac{d{\bm \theta}_0}
{(2\pi)^{N}}\beta\nonumber\\
&=&i\oint_{\mathcal O} \frac{\oint d{\bm \theta}_0}{(2\pi)^{N}}\Big\{
\langle\psi|\frac{\partial}{\partial\bm{\theta}}|\psi\rangle\cdot 
d\bm{\theta}+
\langle\psi|\frac{\partial}{\partial\r}|\psi\rangle\cdot d\r\Big\}\nonumber\\
&=&\bm{I}\cdot\oint_{\mathcal O} d\bm{\theta}+\gamma({\mathcal O})\,,
\label{eq:aa_gb1}
\ea
where we have introduced a new notation $\bar{\beta}$ to stand for the averaged 
total phase $\beta$. We may also call $\bar{\beta}$ the total phase of an 
adiabatically evolving manifold. In the derivation, we have noticed for $\gamma$ 
that the integral over $\bm{\theta}_0$ is the same as over 
a $\bm{\theta}$ at any given $\r$.  We have also 
used that
\be
I_j=\frac{1}{2\pi}\oint d{\theta_0}_j
\langle\psi|\frac{\partial}{\partial\theta_j}
|\psi\rangle\Big|_{\theta_j={\theta_0}_j}
\ee
is the action of the motion associated with angle $\theta_j$.  
Note that this expression is also the Aharonov-Anandan (AA) phase of the 
cyclic state, describing the motion associated with the angle 
variable $\theta_j$\cite{liu}.

The equation (\ref{eq:aa_gb1}) shows that the averaged total phase $\bar{\beta}$ 
has two parts: a dynamical part involving action-angle variables
and a geometric part that is exactly our new phase
$\gamma$. For the special case of eigenstate $|\psi\rangle=|E_n\rangle$, 
since there is only one non-zero action $I=\langle E_n|E_n\rangle$, we have
\be
\bar{\beta}=\beta=\langle E_n|E_n\rangle \int_0^T\omega_n dt+\gamma_n(\mathcal O)\,.
\ee
It recovers the well-known fact that Berry's phase is the geometric 
part of the total phase of an adiabatic evolution of an eigenstate\cite{berry}. 

\begin{figure}[!t]
\vskip10pt
\center{\includegraphics[width=8.0cm]{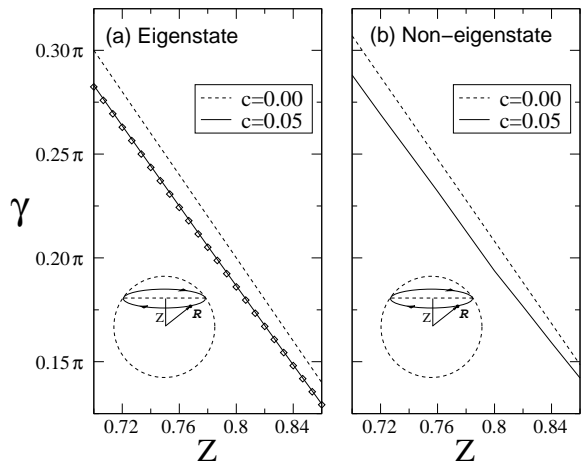}}
\caption{Geometrical phases $\gamma$ of the nonlinear two-level 
model (\ref{eq:gnlz1},\ref{eq:gnlz2}): (a) the lower eigenstate and
(b) the non-eigenstate near the lower eigenstate.
The non-eigenstate has an action (or AA phase) of $I=0.005$.
The insert shows the closed path $\mathcal O$ used 
which is a circle on a unit sphere with $Z$ fixed. 
The circle is traversed with a rate of 0.001. The results
for eigenstates are compared to the analytical expression of 
Eq.(\ref{eq:be2}) (denoted by diamonds): a very good
match is found.}
\label{fig:suph}
\end{figure}

We re-write Eq.(\ref{eq:aa_gb1}) and obtain another
expression for the geometric phase $\gamma$,
\be
\label{eq:aa_gb2}
\gamma({\mathcal O})=\bar{\beta}-\bm{I}\cdot\oint_{\mathcal O}
d\bm{\theta}\,.
\ee
This complicated expression turns out to be easier to be implemented numerically;
we will use it to compute  the results shown in Fig.\ref{fig:suph}. 

One can repeat the derivation from Eq.(\ref{eq:taa}) to 
Eq.(\ref{eq:aa_gb2}) in the projective Hilbert space. In this situation,
the total phase Eq.(\ref{eq:taa}) becomes an AA phase, and the phase $\gamma$
can then be regarded as the geometric part of the AA phase associated
with the parameter space.

There is another interesting angle looking into the new phase $\gamma$.
As already mentioned, the system described in Eq.(\ref{eq:nls}) has a canonical 
classical Hamiltonian structure. With this classical structure,
we can introduce naturally Hannay's angles, a classical concept, 
into quantum systems. Following the expression for Hannay's
angles in Ref.\cite{hannay,semi1,semi2}, 
we find that these angles in our quantum system (\ref{eq:nls})
are related to the phase $\gamma$ by
\be
\label{eq:qhan}
\bm{\alpha}({\mathcal O})=-\frac{\partial}{\partial\bm{I}}\gamma({\mathcal O})\,.
\ee
In  linear quantum systems, these Hannay's angles differ from Berry's phases 
of eigenstates only by a sign, $\alpha_n=-\gamma_n$, 
according to Eq.(\ref{eq:lingb}). Note the relation (\ref{eq:qhan}) is very 
similar in  form to the semiclassical relation
between Hannay's angles and Berry's phases derived in Ref.\cite{semi1,semi2}
although they are two very different relations. 
So far, it is not clear why the similarity\cite{jie}.

Finally, we demonstrate the geometric phase $\gamma$ with
a nonlinear two-level model as given by
\ba
\label{eq:gnlz1}
i\frac{d\varphi_1}{d t}&=&\Big[c|\varphi_2|^2+\frac{Z}{2}\Big]\varphi_1+
\frac{X-iY}{2}\varphi_2 \,,\\
\label{eq:gnlz2}
i\frac{d\varphi_2}{d t}&=&\Big[c|\varphi_1|^2-\frac{Z}{2}\Big]\varphi_2+
\frac{X+iY}{2}\varphi_1\,.
\ea
This simple model can be used to describe the Josephson effect of Bose-Einstein
condensates residing in a double-well potential\cite{leggett,walls}. The complex
coupling constant, as denoted by $X$ and $Y$, can be realized in experiment
through phase imprinting on one of the two wells\cite{imprint}.

We first look at the geometric phase $\gamma$ for eigenstates of 
this nonlinear quantum system. This is to find all the eigenstates 
$|E_n(\r)\rangle$ for a closed path $\mathcal O$, and use them to 
calculate the phase with Eq.(\ref{eq:be}). Both steps could be done 
numerically; fortunately for this simple case, analytical results can 
be obtained. When the path is restricted on the unit sphere $X^2+Y^2+Z^2=1$,
we find that the phases for these nonlinear eigenstates are
\be
\label{eq:be2}
\gamma({\mathcal O})=\int_{\partial S={\mathcal O}}\frac{\eta^3(\r+c\eta\hat{z})\cdot d\bm{S}}
{(c\eta+Z)^2(c\eta^3+Z)}\,,
\ee
where $\hat{z}$ is the unit vector along the $z$-axis and $\eta$ is one of
the real roots of 
\be
\label{eq:proot}
c^2\eta^4+2cZ\eta^3+(1-c^2)\eta^2-2cZ\eta-Z^2=0\,.
\ee
Different real roots $\eta$ correspond to different eigenstates. 
It is clear that Eq.(\ref{eq:proot})
can have more than two real roots, indicating that there can be
more than two eigenstates\cite{nlz}. Here
we limit ourselves to  the situations where Eq.(\ref{eq:proot}) has
only two real roots. 
For the path $\mathcal O$ that is a circle with a fixed $Z$, the geometric
phase in Eq.(\ref{eq:be2}) becomes $\gamma=(1-\eta)\pi$. The diamonds
in Fig.\ref{fig:suph} are calculated with Eq.(\ref{eq:be2}),
showing how $\gamma$ for the lower eigenstate changes with $Z$.

For a general quantum state, we have to resort to numerical means.
The path $\mathcal O$ is picked to be a circle with fixed $Z$.
We then solve Eqs.(\ref{eq:gnlz1},\ref{eq:gnlz2}) numerically after
choosing a changing rate $v=0.001$ for the parameters $\r=\{X,Y,Z\}$.
The evolving states are recorded and used to compute the phase $\gamma$ 
with Eq.(\ref{eq:aa_gb2}), where the averaging is done for
different initial states with the same action (or AA phase) $I$.
Results for $c=0.05$ are plotted in 
Fig.\ref{fig:suph}, showing how the phase changes with $Z$. 
Computation is done for both eigenstate and non-eigenstate and 
the results (solid lines) are compared to the phases for 
the linear case $c=0.0$ (dashed lines). The changing rate of $\r$ ($v=0.001$) 
is slow enough to be considered as adiabatic. This is witnessed by the good 
fit between the solid line and the diamonds in Fig.\ref{fig:suph}(a) 
as the diamonds are the analytical results of Eq.(\ref{eq:be2}).

In summary, we have introduced a geometric phase for a general 
adiabatically evolving quantum state. 
The new phase to certain extent unifies two different concepts,
Berry's phase and Hannay's angles. It is very interesting to find
potential applications for this new geometric phase while its 
properties are being further explored.

We acknowledge the support of NSF (DMR-0306239), 
R. A. Welch Foundation, and LDRD of ORNL, managed by UT-Battelle,
LLC for the USDOE (DE-AC0500OR22725).

\end{document}